\documentclass{emulateapj}

\def\rpp{Reports of Progress in Physics}%
                \newcommand{\kms}{{km\,s$^{-1}~$}}

\slugcomment{Accepted by ApJ, June 19, 2008}

\shorttitle{The non-thermal emission of Cas A}
\shortauthors{Helder and Vink}

\begin{document}

\title{Characterizing the non-thermal emission of Cas A }

\author{E. A. Helder\altaffilmark{1} and J. Vink\altaffilmark{1}}
\affil{Astronomical Institute Utrecht, Utrecht University, P.O. Box 80000, NL-3508 TA Utrecht, The Netherlands}

\begin{abstract}
We report on our analysis of the 1 Ms Chandra observation of the supernova remnant
Cas A in order to localize,  characterize and quantify its non-thermal X-ray emission. 
More specifically, we investigated whether the X-ray synchrotron emission from the inside of the remnant is from the outward shock, but projected toward the inner ring, or from the inner shell. 
We tackle this problem by employing a Lucy-Richardson deconvolution technique and measuring spectral indices in the 4.2-6 keV band.

We show that most of the continuum emission is coming from an inner ring that coincides with the location of the reverse shock. This inner ring includes filaments, whose X-ray emission has been found to be dominated by X-ray synchrotron emission. 
The X-ray emission from these filaments, both at the forward shock and from the inner ring, 
have relatively hard spectra with spectral index $> -3.1$. 
The regions emitting hard X-ray continuum contribute about 54\% of the total X-ray emission in the 4.2-6 keV. This is lower than suggested by extrapolating the hard X-ray spectrum  as measured by BeppoSAX-PDS and INTEGRAL. 
This can be reconciled by assuming a gradual steepening of the spectrum toward higher energies.  
We argue that the X-ray synchrotron emission is mainly coming from the Western part of the reverse shock. The reverse shock in the West is almost at rest in our observation frame, corresponding to a relatively high reverse shock velocity of $\sim 6000$~\kms~in the frame of the freely expanding ejecta.
\end{abstract}

\section{Introduction}

Supernova remnants (SNRs) are the main candidates for producing Galactic cosmic rays, with energies at least up to the so-called knee of the cosmic ray spectrum at $\sim 3 \times 10^{15}$~eV. The first direct evidence for this is the detection of X-ray synchrotron emission caused by $\sim 10^{14}$ eV electrons \citep[first established for SN1006,][]{koyama}. Since the energy of electrons suffers from radiation losses, this might indicate even higher energies for ions. Moreover, hard X-ray tails up to 80 keV have been discovered for several Galactic SNRs \citep{Allen1999}. This has been contributed to either non-thermal bremsstrahlung \citep{Laming2001a} or to synchrotron radiation \citep{Allen1997}. In recent years, additional direct evidence for efficient cosmic ray acceleration has come from detection of TeV $\gamma$-rays for several SNRs by the High Energy Gamma-Ray Astronomy (HEGRA) experiment, the High Energy Spectroscopic System \citep[H.E.S.S, e.g.][]{Aharonian2004} and MAGIC \citep{Albert}. The $\gamma$-ray emission is either caused by inverse Compton scattering by the same electrons that cause X-ray synchrotron emission, or by pion production caused by collisions of accelerated ions with the background plasma.

\begin{figure}[t]
   \centering
   \plotone{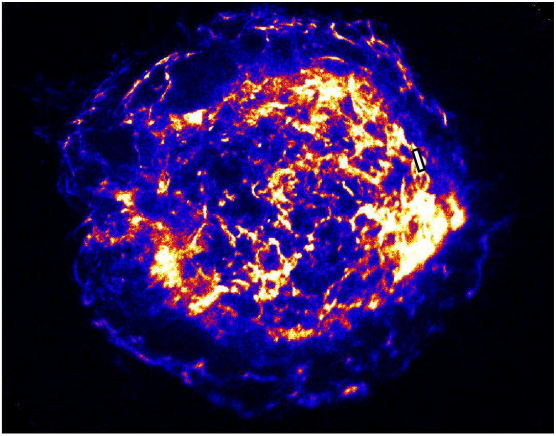}
    \caption{Million second 4-6 keV continuum image of Cas A, obtained with Chandra. The rectangle indicates region `D' in \cite{Hughes}. See the electronic edition of the Journal for a color version 
of this figure.}
              \label{Continuum}%
    \end{figure}

Cassiopeia A (Cas A) is one of the supernova remnants with a hard X-ray tail \citep{The} and has recently also been detected in $\gamma$-rays \citep{Aharonian2001, Albert}.  This remnant was until recently\footnote{Recent expansion measurements of G1.9+0.3 show an age for this remnant of around 100 years \citep{Green, Reynolds2008}.}  the youngest known supernova remnant in the Galaxy; its supernova was probably around 1671 \citep{Thorstensen}. In 2001 Chandra detected thin, X-ray synchrotron emitting, filaments at the forward shock of the remnant \citep[Fig. \ref{Continuum}, see also][] {Gotthelf}. This implies the presence of electrons with energies of $\sim 10^{13}$~eV for the magnetic fields in Cas A, estimated to be 0.1 mG to 0.6 mG \citep{Vink2003, Berezhko}. These synchrotron rims can be understood in the context of diffusive shock acceleration and synchrotron cooling downstream of the shock. 

\begin{figure}[b]
   \centering
     \epsscale{.9}
   \plotone{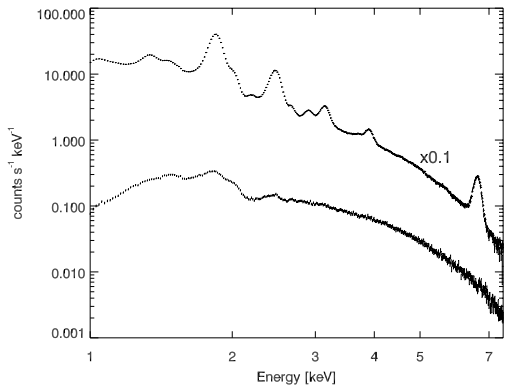}
    \caption{The spectrum of Cas A as observed by Chandra. Below is the spectrum of the featureless filament (`D') described by \cite{Hughes} extracted from the megasecond observation, above is a spectrum of the whole remnant of one single observation (ObsID 4638), multiplied by 0.1.}
              \label{Example}%
    \end{figure}

In addition, the Chandra image shows thin filaments at the inside of the remnant. Some of these inner filaments show a featureless spectrum (Fig. \ref{Example}). \cite{Hughes} identified one of the inner filaments at the West side of the remnant (`region D') as being the projected forward shock, based on its featureless spectrum. \cite{DeLaney} found that the kinematics of the inner filaments, which they interpreted as projected forward shock filaments, are different from the forward shock; they have a lower velocity. 

    \notetoeditor{We would like to have Fig 1. in grayscale in the printed version and in color in the online version }
    
Diffusive shock acceleration is a process which accelerates cosmic rays at a shock front \citep[for a review, see][]{Malkov}. This mechanism accelerates charged particles of sufficient energy, which scatter on turbulent magnetic fields/plasma waves on both sides of the shock front. Each time the shock front is crossed, the particle gains energy, due to the difference in plasma velocity between both sides of the shock front. The higher the difference in the velocities is, the more energy is gained in one iteration and the higher the magnetic field and magnetic field turbulence, the more often particles cross the shock front.  
Since at the location where efficient particle acceleration takes place recently accelerated electrons are present, these locations show X-ray synchrotron radiation. However, further downstream from the shock front, synchrotron losses result in lower maximum energies of the synchrotron radiation.  

The synchrotron spectrum can be approximated over a large range in frequencies with a power-law in flux density: $F_\nu \propto \nu^{-\alpha}$  with an index ($\alpha$) related to the power-law index of the electron distribution ($p$) as: $\alpha = (p-1)/2$. In what follows, we use index $\Gamma$, which refers to the number density index $\Gamma = -(\alpha + 1)$ and $n(E) \propto E^\Gamma$. Near the maximum electron energies, the electron spectrum has an exponential cut-off, but the resulting synchrotron spectrum cuts off less abruptly, roughly as $\exp(-{\sqrt{\nu/\nu_{max}}})$  \citep{Zirakashvili}. In contrast, the other important continuum emission process, thermal bremsstrahlung, has an exponential cut-off ($\propto \exp({-h\nu/kT})$). In Cas A, the plasma temperature ranges between 0.6 and 3.6~keV \citep{Yang}. One should take into account that these temperatures originate from a thermal model. If a partly non-thermal spectrum is fitted with a thermal model, the fitted temperature tends to increase with respect to the real temperature of the plasma. For a thermal bremsstrahlung spectrum with a temperature of 3.5~keV, the power-law index between 4.2 and 6.0 keV is -2.8. For a synchrotron spectrum at the forward shock of Cas A, we typically measure a power-law index of -2.1. We therefore expect the bremssstrahlung continuum to be steeper than the synchrotron continuum in the 4 to 6 keV continuum band. 

In this paper we investigate the shape of the continuum spectrum and its spatial distribution in order to 
address several questions pertaining to the shock acceleration in Cas A:
What is the location of the X-ray synchrotron filaments? What fraction of the overall X-ray continuum is thermal and what fraction is non-thermal? And what are the implications for the hard X-ray emission, above 10~keV. We do this by analyzing the Chandra megasecond observation of Cas A.

 \section{The used data} \label{data}
Chandra observed Cas A for one million second in 2004 from Februari to May \citep{megasecond} with the ACIS S3. For extracting the images, we use the CIAO package, version 3.4. To align the separate pointings, the central compact object is taken as a reference point. We mostly concentrate on the 4.2 to 6 keV energy band, but in addition, we also extracted an image in the line of Si XIII He$\alpha$  from 1.80 to 1.90~keV. We corrected this for continuum emission by subtracting the average of two images next to the Si line: in the 1.63 - 1.64 and 2.13 - 2.14~keV bands. 

We compared the results of our analysis of the Chandra data with a VLA radio map, made 
in the 6 cm band in 2000-2001. This map was kindly provided to us by Tracey DeLaney \citep{DeLaney2005}. The VLA image has a resolution of $0.38\arcsec \times 0.33 \arcsec$, which is comparable to the Chandra telescope resolution of $0.42\arcsec$. However, the Chandra pixels size of $0.49\arcsec$ slightly undersamples the Chandra resolution.

In section \ref{spectra}, we make use of BeppoSAX observations made in June 2001,
with an exposure time of 501 ks. The hard X-ray spectrum obtained with the PDS instrument is described in \cite{Vink2001}. The data obtained with the MECS instruments were never published before, but the details of the analysis are similar to the analysis described in \cite{Vink1999} for a shorter exposure.

Finally, the INTEGRAL-ISGRI spectrum used in section \ref{spectra} is described in \cite{Renaud}.

\begin{figure}[!htb]
\begin{center}

  \epsscale{.9}
   \plotone{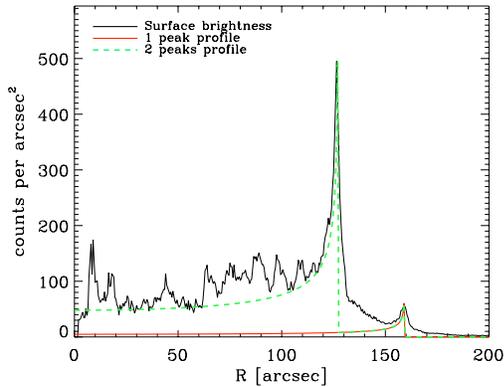}

\caption{Radial surface brightness profile (smooth solid line) in the 4.2 to 6.0~keV energy band at an angle of $10^{\circ}$ to $30^{\circ}$ (including the featureless filament found by \cite{Hughes}). The smooth solid line indicates what the profile will look like if the surface brightness is produced by an emissivity function with only a peak at the outer shock. For the dashed line, we use an emissivity function with two peaks. Note that this is not a fit, just an illustrative example.}
              \label{profile}%
\end{center}
\end{figure}

 \notetoeditor{We would like to have Fig 3. in grayscale in the printed version and in color in the online version }

\section{Separating the forward and reverse shock} \label{Lucy}
The surface brightness ($\Sigma(r)$) of an optically thin object consists of the emissivity function of this object ($\epsilon(r)$), integrated along the line-of-sight. For a spherically symmetric object, this integral is as follows:
\begin{equation}\label{sigma}\Sigma(r) = 2\int ^R _r \epsilon(r')\frac{r'}{\sqrt{r'^2 - r^2}}dr'\end{equation}
In which $R$ denotes the outer radius of this object. 
We do the deconvolution in cylindrical coordinates, with $\theta$ perpendicular to $r$, and $\theta = 0$ is defined in the West, increasing counterclockwise. The number of counts in a sector at inner radius $r$, angular width $d\theta$ (in radians) and thickness $dr$ is:
 \begin{equation}\label{counts}C_r = \Sigma(r)d\theta\left (rdr+\frac{(dr)^2}{2}\right )\end{equation}
By dividing Cas A in 18 sectors of $20^{\circ}$ each, we first make surface brightness profiles of the X-ray continuum between 4.2 and 6~keV. For the surface brightness profiles, we adopt the center of expansion \citep{Thorstensen}: $\alpha = 23^{\rm{h}}23^{\rm{m}}27^{\rm{s}}.77$ and $\delta = 58^{\circ}48\arcmin49\farcs4$ (Equinox J2000) and take step-sizes of 0\farcs5. Furthermore, we assume sperical symmetry for each sector individually. For an example of a surface brightness profile, see the radial surface brightness profile between an angle of $10^{\circ}$ to $30^{\circ}$ in Fig. \ref{profile}. In this figure, we can clearly see the outer shock coming up at $~$160\arcsec, as already found by \cite{Gotthelf}. Using equations \ref{sigma} and \ref{counts}, we now make an emission profile in such a way that it fits the surface brightness of the outer shock (smooth solid line). As the line indicates, this thin, outer shell, can not account for all the surface brightness in the center of the remnant (the line which fits the outer peak is for a small $R$ at least 15 times lower than the surface brightness profile). We also see a peak in the surface brightness at 126\arcsec. We can identify this surface brightness peak with the featureless filament described by \cite{Hughes}. If we now include a second emission peak at 126\arcsec (dashed line), we see that a large part of the surface brightness in the center is covered.

To go from surface brightness to emissivity we use a general de-convolution method described by \cite{Lucy} and previously used on SN1006 by \cite{Willingale}. To test this algorithm, we simulated emission functions, convolved them into a surface brightness profile and add Poisonian noise using the IDL routine `poidev' from the NASA IDL Astronomy User's Library \citep{IDL}, we made them in such a way that the peak of the number of counts in one bin of $20^{\circ}\times 0\farcs5$ is 11000, which corresponds to the peak of Fig. \ref{profile} in terms of counts per $0\farcs5 \times 20^{\circ}$ bin. After constructing this surface brightness profile we used the Lucy algorithm to recover the emissivity function. We stopped the de-convolution after 30 iterations. The resulting emissivity function is similar to the original one, within 10\% for $R>40$\arcsec and even within 5\% for $R>75$\arcsec. 

\subsection{\label{results}Results of the deconvolution}
\begin{figure*}[!t]
\begin{center}
  \epsscale{.9}
   \plotone{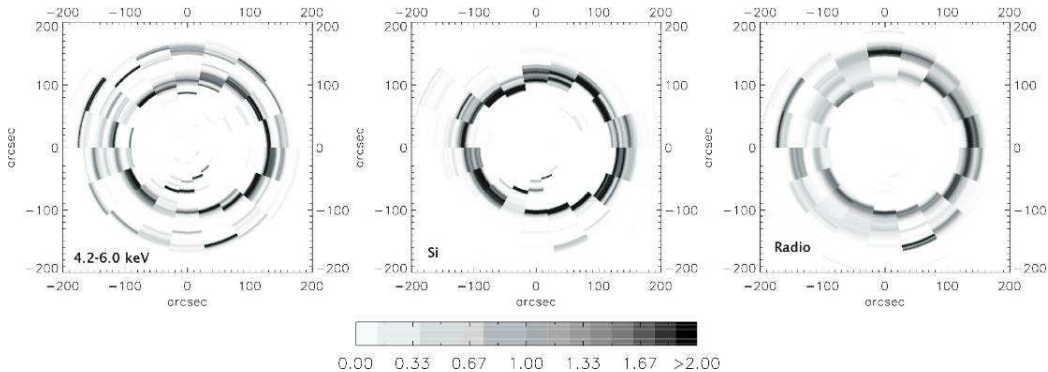}

\caption{\label{diagram}In the left figure we see the de-convolution of the 4.2 to 6.0 keV continuum radiation. For all the figures, the deconvolution is done in sectors, within each sector, spherical symmetry is assumed. The scale is in \% per bin. Each sector is scaled in such a way that the total adds up to 100\%. The middle figure shows the de-convoluted Si image and the right figure the deconvolution of the radio emission. The emissivity is multiplied with the volume of the shell to get the total emission over the whole sphere for each $dr$.  }
\end{center}
\end{figure*}

The results of the de-convolution of the Chandra data of Cas A are given in Fig. \ref{diagram}. We see the forward shock coming up in the X-ray continuum at 160\arcsec and with a width of 10\arcsec. What we also see, is another ring inside the forward shock with a radius of 115\arcsec with a width of 30\arcsec, shifted to the right by 15\arcsec. The forward shock and this inner ring are two almost complete circles, except for the NE region, where the jet is present. If we do the same for the radio and Si images of Cas A, we see a ring at the same location as the inner ring of the X-ray continuum. These inner rings in radio and Si have in previous researches been identified as the reverse shock \citep{Gotthelf}.\\
We note that the results of the deconvolution are far from perfect; the $\chi^2_{red}$ of the individual fits range from 20 to 253 for 4.2 to 6.0 keV and from 18 up to 1150 for the Si band. This is to be expected, since the filamentary structure is clearly inconsistent with the assumed spherical symmetry. Moreover,
the X-ray Doppler maps of Cas A \citep{Willingale2002} also show deviations from spherical symmetry.
However, the fact that we can trace the inner and outer shells consistently from 18 independent deprojected sectors, argues for the veracity of the two rings.

In order to estimate the total contribution of each shell to the total emissivity of Cas A in each band,
we multiply the deprojected shells with the volume of the shells ($4\pi R^2 \Delta R$), taking into
account the relative contributions of each sector.  For the continuum band in the East to South ($\theta \in [180^{\circ}, 260^{\circ}]$), for which the overall X-ray radiation is dominated by
silicon line emission, 16 \% comes from the outer ring and 84 \% comes from the inner ring. For the 
Western part of the remnant ($\theta \in [-100^{\circ}, 20^{\circ}]$), in which most of the filaments dominated by continuum are situated, we find  that 18 \% of the continuum emission is
due to the outer ring and 82\% due to the inner ring.
 
 \section{Thermal Bremsstrahlung versus Synchrotron radiation} \label{spectra}
 
  \begin{figure}
   \centering
     \epsscale{.9}
   \plotone{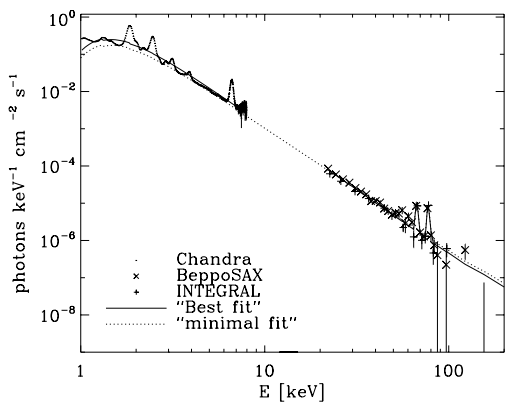}
    \caption{Unfolded broadband X-ray spectrum of Cas A. The solid line represents the best fit for the combined BeppoSAX/PDS and INTEGRAL/ISGRI data. Extrapolation of this line back to the Chandra regime fits the continuum surprisingly well. The dotted line indicates the minimal number of counts in the 4.2 to 6.0 keV band, allowed within the errors of the extrapolation of the hard X-ray fit. The Chandra broadband spectrum of Cas A was made from one of the observations (ObsID 4638) of the megasecond observation \citep{megasecond}. }
              \label{powerlaw}%
    \end{figure}

The hard X-ray emission from Cas A is clearly non-thermal in nature \citep{The, Allen1997, Favata}. In order to see how this relates to the Chandra spectrum of Cas A, we show in Fig. \ref{powerlaw} both the broadband (1-8 keV)  Chandra spectrum and the hard X-ray spectrum (above 20 keV) as obtained with BeppoSAX/PDS \citep{Vink2001} and INTEGRAL/ISGRI \citep{Renaud}. The model shown is fitted to the hard X-ray data only and is similar to the one used by Renaud et al., which includes the contributions of $^{44}$Ti decay at 67.8 and 78.4 keV. We included absorption with  N$_{\rm H} = 1.3\times10^{22}\rm{cm}^{-2}$. Of interest here are the best fit parameters of the non-thermal component, fitted with a power-law. Our best fit parameters are $\Gamma = -3.4 \pm 0.2$ and the normalisation is 3.22 $\pm 1.9$ counts keV$^{-1}$ cm$^{-2}$ s$^{-1}$ at 1 keV.  We checked these results with previous results on the hard X-ray tail of the spectrum of Cas A. \cite{Rothschild} observed Cas A for 226 ks with RXTE and fitted a power-law to the HEXTE data from 20 to 200 keV. They find a power-law index of -3.125 $\pm$ 0.050 with a normalization of $2.0\pm 0.6$ counts cm$^{-2}$s$^{-1}$ at 1 keV. 

We see that  the BeppoSAX power-law, extrapolated back to Chandra energies, fits the continuum between 2.0 - 7.0 keV well. Within a confidence level of 90\%, the extrapolated powerlaw contains a minimum of 93\% of the counts in the Chandra continuum bands.  The results of \cite{Rothschild} amounts to a continuum flux from 4 to 6 keV which is 28\% of the continuum flux measured by Chandra. Note that if we trust the power-law continuum model, there is little room for an additional thermal component. This is surprising, since we know that the continuum in the 1-10 keV range is at least partially due to thermal bremsstrahlung, which must accompany the copious X-ray line emission. In fact, fitting the spectrum in the 0.5-10 keV band can be done with a pure thermal model 
 \citep{Vink1996, Willingale2002}.

To pinpoint the different mechanisms which contribute to the continuum in the Chandra band, we fit a power-law to regions of $4.9\arcsec \times 4.9 \arcsec$ ($10 \times 10$ pixels) for the band between 4.2 and 6.0 keV. We use the eventfiles of the megasecond observation, which we merge, using the central compact object as a reference point. We use one ARF\footnote{Ancillary Response File, which contains a description of the effective area of the instrument.} file, made for the spectrum of Fig. \ref{powerlaw}. We estimated the background contribution, using an annulus around Cas A, with $R_{min}=205\farcs 8$ and $R_{max} = 235\farcs 2$ and a center as defined in section \ref{Lucy}. We neglected the effects of differential absorption over this small band; in the most extreme cases, the absorption varies from $1\times 10^{22}$ to $1.7 \times 10^{22} \rm{cm}^{-2}$ which has an effect of at maximum 0.09 in the power-law index. Fitting the continuum of the total remnant between 4.2 to 6.0 keV, using N$_{\rm H} = 1.3\times 10^{22}\rm{cm}^{-2}$, we found that the average index is -3.14. 

\begin{figure*}[!htb]
\begin{center}
\plotone{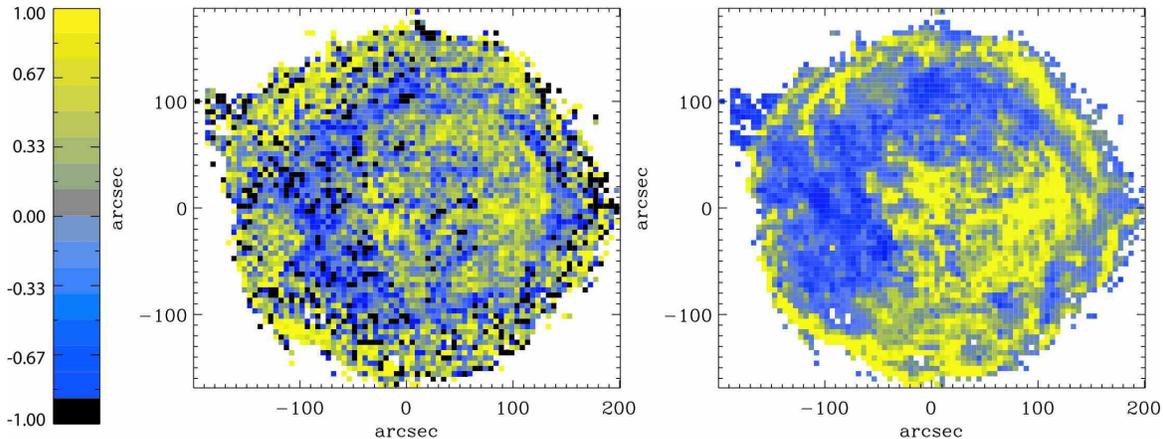}
\caption{In the left Figure, we see the fitted power-law indices ($\Gamma$) + 3.1, to show the difference between the individual fits and the overall spectrum. Dark indicates a steeper spectrum. In the right Figure, we see a map of an image in the 4.2-6.0 keV continuum bands divided by a broadband image. The lighter color means relatively more continuum and thus a harder spectrum. Note the similarities between the images: as already noted by \cite{DeLaney}, the continuum dominated areas tend to have a harder spectrum. Even so, not all hard spectra have a lack of line emission.}
              \label{specsel}%
\end{center}
\end{figure*}

 \begin{figure}
   \centering
     \epsscale{.9}
   \plotone{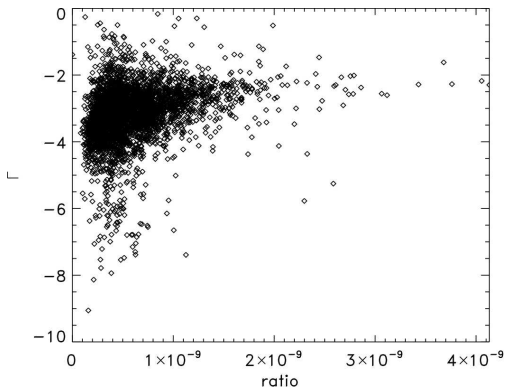}
    \caption{On the x-axis is the ratio of continuum over broadband in arbitrary units (the same units as in Fig. \ref{specsel}, right image), a higher value means relatively more continuum. On the y-axis the fitted powerlaw indices of the corresponding bins.}
              \label{sum}%
    \end{figure}

Fig. \ref{specsel} shows that the power-law index varies considerably inside Cas A. We see that some of the regions with hard spectra overlap with known regions of non-thermal X-ray emission. For example the forward shock region appears to have a harder power-law. And also the non-thermal filaments in the West show up in the spectral index map. However, there are also is also hard continuum emission in regions where there is also line emission. This is illustrated by Figures \ref{specsel} and \ref{sum}.

\begin{figure}[!htb]
\begin{center}
  \epsscale{.9}
   \plotone{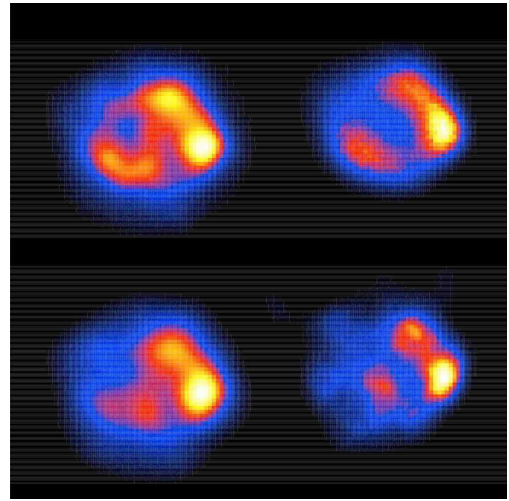}
\caption{The upper row is a Chandra image of Cas A in the 4 to 6 keV band, convolved to the BeppoSAX resolution (left) and the BeppoSAX image of Cas A in the same energy band (right). The second row has on the left side the image of Fig. \ref{specsel}, extrapolated to 9 to 11 keV (Fig. \ref{911extraChandra}) and convolved to the BeppoSAX resolution. The right image shows the actual BeppoSAX image in the 9 to 11 keV band. The extrapolation matches qualitatively the BeppoSAX image and therefore, the powerlaw fits, give a reasonable prediction of the hard X-ray image of Cas A. }
              \label{911extra}%
\end{center}
\end{figure}

\begin{figure}
\begin{center}
\epsscale{.8}
\plotone{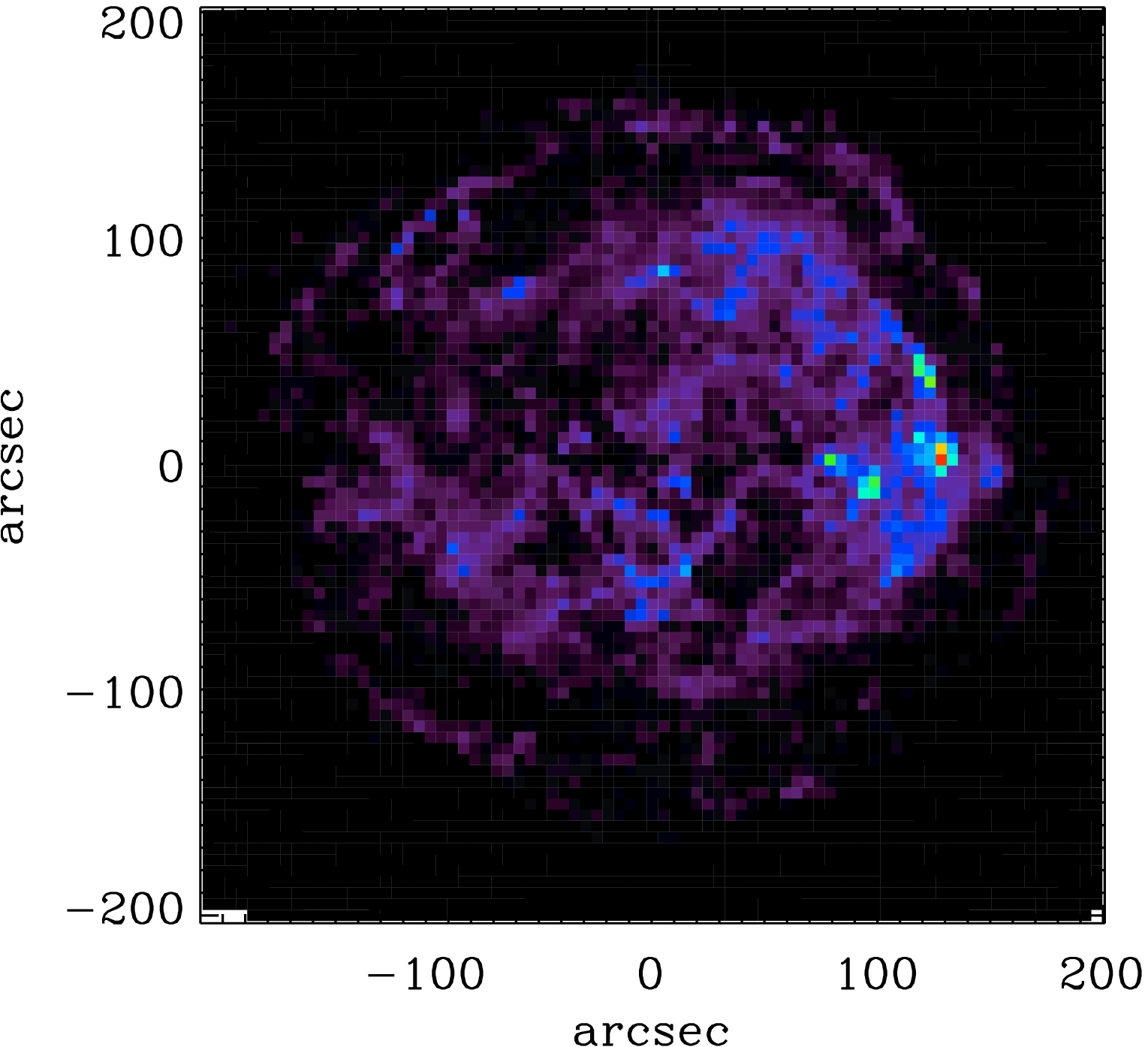}
\caption{An extapolation of the best fit power-laws from Fig. \ref{specsel} to the 9 to 11 keV band }
              \label{911extraChandra}%
\end{center}
\end{figure}

The spectral map of Fig \ref{specsel} implies that if we observe Cas A at higher energies, the Western part of the remnant will become more prominent. We validated this by comparing an extrapolation of our best fit power-law spectra with the 501 ks BeppoSAX/MECS observation in the 9-11 keV band.\footnote{There is also an observation made by XMM-Newton of Cas A up to 15 keV \citep{Bleeker}, because of errors in the exposure map of this observation, we choose to use the BeppoSAX data.} We deconvolved the BeppoSAX image using the Lucy deconvolution method, following the procedure described in \cite{Vink1999}. The results are shown in Fig \ref{911extra}. for comparison, we also show the 4 to 6 keV BeppoSAX image next to the Chandra image smoothed to roughly the same resolution (the extrapolation without smoothing is shown in Fig \ref{911extraChandra}). Qualitatively the 9-11 keV image of BeppoSAX agrees with the extrapolated Chandra image: the South-Eastern part of the remnant  is relatively less bright, whereas the Western part and the Southern part of the center are becoming more prominent.

\begin{figure}
   \centering
   \epsscale{.8}
   \plotone{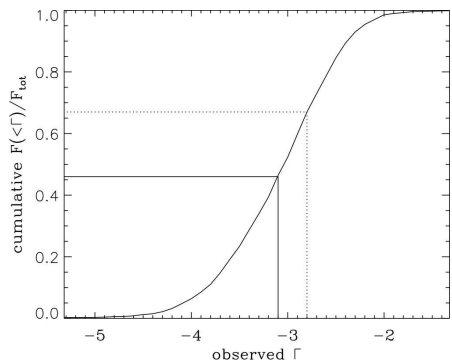}
    \caption{ A plot of the cumulative flux between 4.2 and 6 keV versus the best-fit powerlaw index as found in Fig \ref{specsel}. The solid line shows that 46\% of the flux has a power-law index steeper than -3.1. The dotted line shows that 33\% of the flux has a power-law flatter than -2.8; the power-law index associated with a thermal bremsstrahlung model with kT = 3.5 keV.  }
              \label{integrated}%
    \end{figure}

Since for a bremsstrahlung continuum, we expect an exponential cut-off and thus a soft spectrum and for synchrotron radiation, we expect a hard spectrum, we tentatively identify the hard spectra with synchrotron emission. Although there is a likely overlap in spectral index between thermal and non-thermal emission we can nevertheless estimate the total contribution of non-thermal emission by noting that in Fig. \ref{specsel} left, those regions that have abundant line emission (the Eastern part of the shell), have power-law spectra steeper than -3.1. The total flux associated with those power-law indices amounts to 46 \% of the total flux in the 4.2 to 6 keV band (Fig. \ref{integrated}). Suggesting that the other 
54\% of the flux, is due to non-thermal radiation. This corresponds roughly to a total non-thermal flux above 4  keV of 2.7 $\times 10^{-10}{\rm erg~s^{-1}~cm^{-2}}$. If we take a power-law index of -2.8 as an upper limit, this corresponds to a thermal bremsstrahlung model with kT = 3.5 keV, the non-thermal contribution comes down to 33 \%. This is indeed a large fraction of the total contintuum emission, but not as large as the 93\% suggested by extrapolation of the hard X-ray power-law. The reason for this is probably that the true spectral shapes are not exactly power-laws, but steepen at higher energies. So apparently, in the Chandra band the addition of soft thermal emission and hard non-thermal gives by coincidence  almost the same power law index, as the steepening non-thermal spectra at high energies.
The lack of an obvious spectral break is probably due to the variation in power law indices for the non-thermal spectra, and the variation in cut-off energies across the remnant.

\subsection{\label{West}West side of Cas A}
So far, we have shown that a large part of the non-thermal emission is associated with the reverse shock region and that most of this emission is coming the Western part of the remnant. This prompted us to investigate the spectrum of the Western part more closely. Specifically we were interested in whether the  reverse shock in the Western part shocks the less dense material. The reason is that the Western region not only shows more non-thermal emission, but there also seems to be less thermal emission. A low density of a shocked material results, apart from less flux from shock-heated plasma, also in   a low ionization time scale, $n_{\rm e}t$. With this in mind, we extracted spectra from one of the featureless filaments, located at the reverse shock. This is the same filament as used by \cite{Hughes} (filament `D'). We extracted the spectrum of this region in the individual eventfiles of the megasecond image and added them with the \texttt{addspec} tool of ftools in heasoft. We fit the resulting spectrum with a model consisting of a power-law and the non-equilibrium ionization (NEI) model of SPEX \citep{spex}. The best fit parameters are listed in Table \ref{nei}.

\begin{deluxetable}{cc}
\tablecolumns{8}
\tablewidth{0pc}
\tablecaption{\label{nei} Results of the spectral fit to the featureless filament}
\tablehead{
\colhead{Parameter} & \colhead{Value}  }
\startdata
 n$_{\rm{H}}$ (cm$^{-2}$)&  1.67$^{+0.03}_{-0.03}\times 10^{22}$ \\
n$_{\rm{e}}$t (cm$^{-3}$ s) &  8.6$^{+1.0}_{-1.0}\times 10^{10}$\\
 kT (keV) &  1.8$^{+0.15}_{-0.15}$\\
$\Gamma$ & -2.09$^{+0.07}_{-0.07}$ \\
$\rm{F_{pow}}/\rm{F_{nei}}$ ($E \in [0.4~\rm{keV},8~\rm{keV} ]$)& 2.23\\
$\chi^2$/d.o.f & 283/145\\

\enddata
\tablecomments{Previous studies \citep{Keohanenh, Willingale2002} show that the absorption in the West side is higher than in the East, with values up to 1.7 $\times 10^{22}~\rm{cm^{-2}}$. So a value of 1.67 $\times 10^{22}~\rm{cm^{-2}}$ is still within the expectations.}
\end{deluxetable}

\section{Discussion} \label{discussion}
Using the a combination of deprojection and characterization of the X-ray continuum power-law slope of the one megasecond Chandra observation, we have determined the location of the non-thermal X-ray emission of Cas A. We have found that a significant fraction of the X-ray continuum emission is coming from two separate rings; one can be identified with the forward shock, the other with the shell dominated by bright ejecta. About 54\% of all continuum emission in the 4.2 to 6 keV band is likely to be non-thermal. 

The non-thermal emission from the forward shock is likely due to synchrotron radiation from electrons accelerated at the forward shock, confined to the forward shock region by synchrotron cooling \citep{Vink2003, Berezhko}.  However, the non-thermal emission of Cas A as a whole is dominated by the contribution of the inner ring. This inner region of emission is shifted to the West of the remnant by $\sim15\arcsec$. A similar shift was found for the location of the reverse shock by \cite{Reed, Gotthelf}, on the basis of optical data and a deprojection of radio data and a 50 ks Chandra observation, in the Si band. \cite{Gotthelf} also found marginal evidence for a higher emissivity at the reverse shock in the 4 to 6 keV band using the 50 ks Chandra observation.

\subsection{Nature and location of the inner non-thermal emission}
Since some of the  spectra in the West hardly show any line emission, we think it is likely that this is synchrotron emission, since non-thermal bremsstrahlung without line emission is hard to establish,
unless one has peculiar abundances. Such a bremsstrahlung model was once invoked for SN1006 \citep{Hamilton1986},
but has now been abandoned in favor of synchrotron emission. Furthermore, a recent paper by \cite{Vink2008} shows that electrons with energies close to the thermal electron energy distribution, lose their energy relatively fast due to Coulomb losses. This process is already important in the Chandra energy band for $n_{\rm e}t  < 8.6 \times 10^{10} {\rm cm}^{-3}{\rm s}$; the value reported here for a filament, dominated by continuum emission in the West (section \ref{West}, Table \ref{nei}).  For  $n_{\rm e}t  \sim  10^{11} {\rm cm}^{-3}{\rm s}$, typical for Cas A \citep{Willingale2002}, one only expects to see non-thermal bremsstrahlung for photon energies $\gtrsim 100~\rm{keV}$.

When it comes to the location of the synchrotron emission, there are two possibilities: the reverse shock and the contact discontinuity. The contact discontinuity marks the border between shocked ejecta and shocked circumstellar medium. Hydrodynamical solutions show that the density and the magnetic field peak at this radius \citep{Chevalier1982, Lyutikov}. For a supernova remnant evolving in a stellar wind, the contact discontinuity is close to the reverse shock. So from our deprojections it difficult to judge whether the X-ray synchrotron emission is coming from the reverse shock or the contact discontinuity. It is unlikely that electrons are accelerated to TeV energies at the contact discontinuity, since no viable acceleration mechanism is known \citep[however, see][]{Lyutikov}, but there are two ways of generating X-ray synchrotron emission at the contact discontinuity: 1) Due to an increase of the magnetic field, electrons with relatively low energies suddenly light up in X-rays. 2) High energy electrons and positrons are created through the decay of charged pions ($\pi^\pm$), caused by hadronic cosmic ray collisions
\citep{Gaisser1990}.
Charged pions decay into muons and muon neutrinos. The muons decay into electrons/positrons and electron and muon neutrinos. The electrons and positrons thus created are often called secondary electrons and positrons. 

For option 1: For synchrotron radiation the relation between electron energy and photon energy is 
\begin{equation}
E_{{\rm ph}} = 19 E_{{\rm TeV}}^2B_{\rm G}~{\rm keV} \label{Eph}
\end{equation} 
\citep{Ginzburg1965}. For electrons to be invisible in X-ray synchrotron radiation, the peak energy should be an order of magnitude lower than the X-ray continuum at 4-6 keV. Therefore, the magnetic field should increase an order of magnitude at the contact discontinuity. As an example, an electron, accelerated at the forward shock, with an energy of 3 TeV in the typical magnetic field of Cas A of 0.5~mG, typically emits photons of 0.09~keV. When this electron suddenly enters a region of 5 mG, it will emit photons of typically 1 keV; detectable in X-rays. However, the synchrotron loss time for such an electron in a 0.5~mG magnetic field, 18 years, is rather short compared to the age of Cas A. In this time the particle has to diffuse from the forward shock region to the contact discontinuity, which is about 0.5 pc, for a distance to Cas A of 3.4 kpc \citep{Reed}.   For 3 TeV and B=0.5 mG the diffusion constant is $D=\eta\cdot 2\times 10^{23}~{\rm cm^2s^{-1}}$, with $\eta=1$ corresponding to Bohm diffusion \citep{Malkov}. In 18 yr a particle can diffuse by $R \approx \sqrt{2Dt} = 5\times 10^{-3} \sqrt{\eta}$~pc. In other words, this
model only works if the magnetic field turbulence is very low, corresponding to $\eta > 10000$. This unlikely, since the ample presence of  cosmic rays results in magnetic field turbulence. Moreover, at the shock front $\eta \sim 1$ \citep{Vinkreview2006,Stage2006}.  For low values of $\eta$, the diffusion length ($l_{\rm diff}$ the length for which advection dominates over diffusion), is short: $l_{\rm diff} = \eta 4.5\times10^{-4} $ parsec. So, for low $\eta$, the advection velocity, ($u = v_{\rm{fs}}/4$, with $v_{\rm{fs}}$ the forward shock velocity) is the relevant velocity. For a $v_{\rm{fs}}$ of 5800 \kms, $u=1450$ \kms. Using this velocity, it takes 674 years to go from the forward shock to the contact discontinuity, longer than the synchrotron loss time of 18 years.

It has been argued that the magnetic field at the shock front is high, but rapidly decays towards the inside \citep{Pohl}, increasing the synchrotron loss times, and increasing the diffusion constant. However, the decay in magnetic field should be reflected in the radio emissivity from the forward shock to the inside, which is contrary to observations, that show gradual increase in emissivity toward the center starting at the shock front \citep{Gotthelf}.

For option 2 (secondary electrons): if the X-ray synchrotron emission from the inside is due to secondary electrons, this would be an important discovery. It would be evidence for the presence of TeV ion cosmic rays. Neutral pions ($\pi^0$) are made in comparable quantities to charged pions. The power in secondary electrons should therefore be comparable to the luminosity in pion decay. In order to
compare the fluxes from X-ray synchrotron and $\gamma$-ray radiation one should take into account the conversion from pion energy ($\pi^0$ \& $\pi^\pm$)
to photon energies, both in the TeV band (through $\pi^0$ decay) and in the X-ray band, due to
synchrotron radiation from secondary electrons. Taking into account these various decay channels, one finds that 
$E_{{\rm ph}} \sim 20 E_{{\gamma \rm TeV}}^2B_{\rm G}~{\rm keV}$ \footnote{
The decay product of $\pi^0$\ decay is two photons, with
each, on average, an energy $E_{\gamma \rm TeV} = 0.5E_{\pi^0}$. For the charged pions, $\pi^{\pm}$, the final decay product consists of electrons (positrons) and neutrinos
( $\pi^\pm \rightarrow \mu^\pm + \nu_\mu(\bar{\nu}_\mu)$, $ \mu^{\pm}\rightarrow e^\pm + \nu_e(\bar{\nu}_e) + \nu_\mu(\bar{\nu}_\mu)$). In this case the final
electron or positron also take up, on average, half the initial pion energy $E_{e^\pm} \approx 0.5 E_{\pi^\pm}$; hence, for a given pion energy, $E_{e^\pm}\approx E_{\gamma \rm TeV}$. This electron or positron emits synchrotron
radiation at a peak frequency given by Eq.~\ref{Eph}. Hence the close resemblance of this equation with Eq.~\ref{Eph}.}, so the X-ray synchrotron flux above 4~keV
should correspond to roughly the $\gamma$-ray flux above 20~TeV.
We found for the X-ray synchrotron flux above 4 keV 2.7$\times 10^{-10}$~erg~s$^{-1}$cm$^{-2}$. This is a factor 130 higher than the $\gamma$-ray flux above 1~TeV, which we calculated to be $2.1\times10^{-12}~\rm{erg}~\rm{cm}^{-2}~\rm{s}^{-1}$ using the photon-flux and photon index above 1~TeV reported by the MAGIC collaboration \citep{Albert}. 
Since we should actually have evaluated the flux above 20~TeV, we can rule out that the
X-ray synchrotron emission is caused by secondary electrons.

In our view it is therefore most likely that the X-ray synchrotron emission is caused by
electrons accelerated at the reverse shock. Similarly to the forward shock region the electrons
are likely to be confined to a region near the shock itself.
For a long time the reverse shock as location for acceleration has been neglected, because of
its putative low magnetic field. Moreover, the abundance pattern of cosmic rays is consistent
with acceleration from plasmas with solar abundances \citep{Hoerandel}.  
As far as the magnetic field is concerned, however, 
\citet{ellison} have argued that if magnetic field
amplification \citep{bell} works at the forward shock it is likely to operate at the reverse shock as well.
Also for the SNR RCW 86 it has been suggested that the X-ray synchrotron emission in the
Southwest of the remnant is coming from electrons accelerated at the reverse shock \citep{rho}.

\subsection{Reverse shock velocity in the West and the presence of synchrotron emission}
The presence of X-ray synchrotron radiation from shock accelerated electrons is only expected
if the shock velocity is high enough: shock acceleration theory predicts for the maximum photon energies \citep{Aharonian1999}:
\begin{equation}
E_{\rm ph} = 0.5 \eta^{-1} \bigl(  \frac{v_{\rm s}}{2000{\ \rm km s^{-1}}}\bigr)^2 \ {\rm keV},\label{eq-emax}
\end{equation}
with $v_s$ the shock velocity. Note that the photon energy is independent of the magnetic field.
For the reverse  shock the velocity in Eq.~\ref{eq-emax} refers to the shock speed in the frame of
the ejecta. The ejecta velocity is equal to the free expansion velocity $v_{\rm f, ej} =r_{\rm ej}/t$.
For the reverse shock the shock velocity as seen by the ejecta is therefore
$v_{\rm s, ej}  = v_{\rm f, ej} - v_{\rm s, obs}$, with $v_{\rm s, obs}$ the shock velocity in the
frame of the observer.
The presence of X-ray synchrotron radiation from the reverse shock in the Western half of Cas A suggests a higher reverse shock  than in the rest of the remnant. 
Proper motions of  knots at the inside of Cas A in X-rays were most recently measured by \cite{DeLaney}, but some details
of the measurements, including measurements of the proper motions as a function of azimuth, only appeared in appendix 4.4 of  \citet[][Fig. 4.6]{DeLaneyPhD}.
In this Figure, 
we see that in the West the expansion rate is between -0.1\%yr$^{-1}$ and 0.1\%$^{-1}$ this corresponds
to $v_{\rm{r,ob}}$ at the reverse shock of approximately -2000~\kms to 2000~\kms, implying
a shock velocity \footnote{We have checked this result using the Chandra 1 Ms observation and an observation of Chandra made in 2000
using the procedure described in \citet{Vink1998}. We concentrated on filament 'D'.
The result confirms the lower expansion or even backward velocities:
 we found $-970\pm 140$~\kms, i.e. the filament seems to move towards the center.
This implies a $v_{\rm{s,ej}} \approx  6900~$\kms. \citet{Vink1998} already reported
a lower expansion for the whole Western part of Cas A.}
  $v_{\rm{s,ej}} \approx 3900 - 7900~$\kms. In the Northern and Eastern part of Cas A, \citet{DeLaneyPhD} finds expansion rates of 0.2\%yr$^{-1}$,
corresponding to $v_{\rm{r,ob}} \approx 4000$~\kms and a shock velocity of 
$v_{\rm{s,ej}} \approx 1900~$\kms at the reverse shock.

In the optical \citet{Morse}  reports a shock velocity of  $v_{\rm{s,ej}} \approx 3000~$\kms
for a filament in the Northwest. In our maps (Fig.~\ref{specsel} left) we find that this region has a hard power
law slope, suggesting the presence of synchrotron radiation, but the overall X-ray emission
is dominated by thermal X-ray line emission (Fig.~\ref{specsel} right).
 
 It indeed looks like the presence of X-ray synchrotron emission from the inner
 ring corresponds with reverse shock velocities in excess of 2000~\kms. To some extend
 this is surprising, since the X-ray synchrotron emissivity function is rather broad and 
 can result in some synchrotron emission beyond the maximum photon energy as defined
 in Eq.~\ref{eq-emax}. On the other hand, $\eta = 1$ corresponds to Bohm diffusion, and represents
 the case for maximum acceleration efficiency. The fact that 2000~\kms seems close to
 the velocity dividing the presence or absence of X-ray synchrotron radiation again suggests that
 acceleration takes place close to the Bohm limit. Equation \ref{eq-emax} suggests that the spectra in the West of the remnant will be harder than in the rest, where $v_{\rm{s,ej}}$ is lower.  
  
Note that the high reverse shock velocities in the West (and low observed velocities) are
not in agreement with analytic hydrodynamic solutions for an SNR evolving in a stellar wind 
\citep[][]{Laming}. Adapting the parameters of this model such that the forward shock radius
and velocity and the reverse shock radius, match those of Cas A, we find
 $v_{\rm s, ej}$ ranging between 1600-2300 \kms, significantly lower than in the West, but
 in agreement with those in the Eastern part of Cas A. This suggests that the circumstellar
 density structure is more complex in the West.

\section{Conclusions} \label{Conclusions}
We have presented an anaysis of the spatial and spectral variation of the
X-ray continuum emission of Cas A, based on the 1 Ms Chandra observation.
We find that harder continuum spectra are associated with the filaments, dominated by continuum emission, suggesting that the harder spectra are caused by non-thermal radiation.
A dominant fraction of the non-thermal emission appears to come from the reverse
shock region. We have discussed various options for the nature of the non-thermal
emission and its origins. Some of our conclusions were independently also obtained by \cite{Uchiyama}, but based on X-ray variability. Based on our analysis and discussion we come to the following
conclusions:

\begin{itemize}
\item[$\bullet$] The power-law index of the spectrum between 4.2-6.0 keV is an indicator for X-ray synchrotron emission: there is a correlation between filaments, dominated by continuum emission and hard spectra,
\item[$\bullet$] hard X-ray spectra are not exclusively associated with filaments, dominated by continuum emission, suggesting that non-thermal emission comes also from other regions,
\item[$\bullet$] the non-thermal X-ray emission is likely to be synchrotron radiation,
\item[$\bullet$] the non-thermal accounts for about 54\% of the overall continuum emission in the 4-6 keV band,
\item[$\bullet$] in the Western part of Cas A, most X-ray synchrotron comes from the reverse shock,
\item[$\bullet$] the dominance of X-ray synchrotron emission from the West is probably the result of a locally
higher reverse shock velocity of $v_s \sim 6000$~\kms (corresponding to a lower proper motion) than in the Eastern region ($v_s \sim1900~$\kms).
\end{itemize}

\acknowledgments

We want to thank Tracey DeLaney for providing us with a recent radio map of Cas A.  We would like to thank Frank Verbunt for carefully reading the manuscript and Michele Cappellari for the software we used to display some of our results. This work is supported by the NWO-VIDI grant of J.V. 

{\it Facilities:} \facility{CXO (ACIS)}.


\begin{thebibliography}{54}
\expandafter\ifx\csname natexlab\endcsname\relax\def\natexlab#1{#1}\fi

\bibitem[{{Aharonian} {et~al.}(2001)}]{Aharonian2001}
{Aharonian}, F. {et~al.} 2001, \aap, 370, 112

\bibitem[{{Aharonian} \& {Atoyan}(1999)}]{Aharonian1999}
{Aharonian}, F.~A. \& {Atoyan}, A.~M. 1999, \aap, 351, 330

\bibitem[{{Aharonian} {et~al.}(2004)}]{Aharonian2004}
{Aharonian}, F.~A. {et~al.} 2004, \nat, 432, 75

\bibitem[{{Albert} {et~al.}(2007)}]{Albert}
{Albert}, J. {et~al.} 2007, \aap, 474, 937

\bibitem[{{Allen} {et~al.}(1999){Allen}, {Gotthelf}, \& {Petre,
  R.}}]{Allen1999}
{Allen}, G., {Gotthelf}, E.~V., \& {Petre, R.} 1999, in International Cosmic
  Ray Conference, Vol.~3, International Cosmic Ray Conference, 480--+

\bibitem[{{Allen} {et~al.}(1997)}]{Allen1997}
{Allen}, G.~E. {et~al.} 1997, \apjl, 487, L97+

\bibitem[{{Bell} \& {Lucek}(2001)}]{bell}
{Bell}, A.~R. \& {Lucek}, S.~G. 2001, \mnras, 321, 433

\bibitem[{{Berezhko} \& {V{\"o}lk}(2004)}]{Berezhko}
{Berezhko}, E.~G. \& {V{\"o}lk}, H.~J. 2004, \aap, 419, L27

\bibitem[{{Bleeker} {et~al.}(2001)}]{Bleeker}
{Bleeker}, J.~A.~M. {et~al.} 2001, \aap, 365, L225

\bibitem[{{Chevalier}(1982)}]{Chevalier1982}
{Chevalier}, R.~A. 1982, \apj, 258, 790

\bibitem[{{DeLaney} {et~al.}(2004)}]{DeLaney}
{DeLaney}, T. {et~al.} 2004, \apj, 613, 343

\bibitem[{{Delaney} {et~al.}(2005)}]{DeLaney2005}
{Delaney}, T. {et~al.} 2005, in X-Ray and Radio Connections (eds. L.O.
  Sjouwerman and K.K Dyer) Published electronically by NRAO,
  http://www.aoc.nrao.edu/events/xraydio Held 3-6 February 2004 in Santa Fe,
  New Mexico, USA, (E4.05) 7 pages, ed. L.~O. {Sjouwerman} \& K.~K. {Dyer}

\bibitem[{{Delaney}(2004)}]{DeLaneyPhD}
{Delaney}, T.~A. 2004, PhD thesis, AA(UNIVERSITY OF MINNESOTA)

\bibitem[{{Ellison} {et~al.}(2005){Ellison}, {Decourchelle}, \&
  {Ballet}}]{ellison}
{Ellison}, D.~C., {Decourchelle}, A., \& {Ballet}, J. 2005, \aap, 429, 569

\bibitem[{{Favata} {et~al.}(1997)}]{Favata}
{Favata}, F. {et~al.} 1997, \aap, 324, L49

\bibitem[{{Gaisser}(1990)}]{Gaisser1990}
{Gaisser}, T.~K. 1990, {Cosmic rays and particle physics} (Cambridge and New
  York, Cambridge University Press, 1990, 292 p.)

\bibitem[{{Ginzburg} \& {Syrovatskii}(1965)}]{Ginzburg1965}
{Ginzburg}, V.~L. \& {Syrovatskii}, S.~I. 1965, \araa, 3, 297

\bibitem[{{Gotthelf} {et~al.}(2001)}]{Gotthelf}
{Gotthelf}, E.~V. {et~al.} 2001, \apjl, 552, L39

\bibitem[{{Green} {et~al.}(2008){Green}, {Reynolds}, {Borkowski}, {Hwang},
  {Harrus}, \& {Petre}}]{Green}
{Green}, D.~A., {Reynolds}, S.~P., {Borkowski}, K.~J., {Hwang}, U., {Harrus},
  I., \& {Petre}, R. 2008, \mnras, 387, L54

\bibitem[{{Hamilton} {et~al.}(1986){Hamilton}, {Sarazin}, \&
  {Szymkowiak}}]{Hamilton1986}
{Hamilton}, A.~J.~S., {Sarazin}, C.~L., \& {Szymkowiak}, A.~E. 1986, \apj, 300,
  698

\bibitem[{{H{\"o}randel}(2008)}]{Hoerandel}
{H{\"o}randel}, J.~R. 2008, Advances in Space Research, 41, 442

\bibitem[{{Hughes} {et~al.}(2000){Hughes}, {Rakowski}, {Burrows}, \&
  {Slane}}]{Hughes}
{Hughes}, J.~P., {Rakowski}, C.~E., {Burrows}, D.~N., \& {Slane}, P.~O. 2000,
  \apjl, 528, L109

\bibitem[{{Hwang} {et~al.}(2004)}]{megasecond}
{Hwang}, U. {et~al.} 2004, \apjl, 615, L117

\bibitem[{{Kaastra} {et~al.}(1996){Kaastra}, {Mewe}, \&
  {Nieuwenhuijzen}}]{spex}
{Kaastra}, J.~S., {Mewe}, R., \& {Nieuwenhuijzen}, H. 1996, in UV and X-ray
  Spectroscopy of Astrophysical and Laboratory Plasmas : Proceedings of the
  Eleventh Colloquium on UV and X-ray ... held on May 29-June 2, 1995, Nagoya,
  Japan. Edited by K. Yamashita and T. Watanabe. Tokyo : Universal Academy
  Press, 1996. (Frontiers science series ; no. 15)., p.411, ed. K.~{Yamashita}
  \& T.~{Watanabe}, 411--+

\bibitem[{{Keohane}(1996)}]{Keohanenh}
{Keohane}, J.~W. 1996, in Astronomical Society of the Pacific Conference
  Series, Vol.~99, Cosmic Abundances, ed. S.~S. {Holt} \& G.~{Sonneborn},
  362--365

\bibitem[{{Koyama} {et~al.}(1995)}]{koyama}
{Koyama}, K. {et~al.} 1995, \nat, 378, 255

\bibitem[{{Laming}(2001)}]{Laming2001a}
{Laming}, J.~M. 2001, \apj, 546, 1149

\bibitem[{{Laming} \& {Hwang}(2003)}]{Laming}
{Laming}, J.~M. \& {Hwang}, U. 2003, \apj, 597, 347

\bibitem[{{Landsman}(1993)}]{IDL}
{Landsman}, W.~B. 1993, in Astronomical Society of the Pacific Conference
  Series, Vol.~52, Astronomical Data Analysis Software and Systems II, ed.
  R.~J. {Hanisch}, R.~J.~V. {Brissenden}, \& J.~{Barnes}, 246--+

\bibitem[{{Lucy}(1974)}]{Lucy}
{Lucy}, L.~B. 1974, \aj, 79, 745

\bibitem[{{Lyutikov} \& {Pohl}(2004)}]{Lyutikov}
{Lyutikov}, M. \& {Pohl}, M. 2004, \apj, 609, 785

\bibitem[{{Malkov} \& {Drury}(2001)}]{Malkov}
{Malkov}, M.~A. \& {Drury}, L. 2001, \rpp, 64, 429

\bibitem[{{Morse} {et~al.}(2004)}]{Morse}
{Morse}, J.~A. {et~al.} 2004, \apj, 614, 727

\bibitem[{{Pohl} {et~al.}(2005){Pohl}, {Yan}, \& {Lazarian}}]{Pohl}
{Pohl}, M., {Yan}, H., \& {Lazarian}, A. 2005, \apjl, 626, L101

\bibitem[{{Reed} {et~al.}(1995){Reed}, {Hester}, {Fabian}, \& {Winkler}}]{Reed}
{Reed}, J.~E., {Hester}, J.~J., {Fabian}, A.~C., \& {Winkler}, P.~F. 1995,
  \apj, 440, 706

\bibitem[{{Renaud} {et~al.}(2006)}]{Renaud}
{Renaud}, M. {et~al.} 2006, \apjl, 647, L41

\bibitem[{{Reynolds} {et~al.}(2008){Reynolds}, {Borkowski}, {Green}, {Hwang},
  {Harrus}, \& {Petre}}]{Reynolds2008}
{Reynolds}, S.~P., {Borkowski}, K.~J., {Green}, D.~A., {Hwang}, U., {Harrus},
  I., \& {Petre}, R. 2008, \apjl, 680, L41

\bibitem[{{Rho} {et~al.}(2002){Rho}, {Dyer}, {Borkowski}, \& {Reynolds}}]{rho}
{Rho}, J., {Dyer}, K.~K., {Borkowski}, K.~J., \& {Reynolds}, S.~P. 2002, \apj,
  581, 1116

\bibitem[{{Rothschild} \& {Lingenfelter}(2003)}]{Rothschild}
{Rothschild}, R.~E. \& {Lingenfelter}, R.~E. 2003, \apj, 582, 257

\bibitem[{{Stage} {et~al.}(2006){Stage}, {Allen}, {Houck}, \&
  {Davis}}]{Stage2006}
{Stage}, M.~D., {Allen}, G.~E., {Houck}, J.~C., \& {Davis}, J.~E. 2006, Nature
  Physics, 2, 614

\bibitem[{{The} {et~al.}(1996)}]{The}
{The}, L.-S. {et~al.} 1996, \aaps, 120, C357+

\bibitem[{{Thorstensen} {et~al.}(2001){Thorstensen}, {Fesen}, \& {van den
  Bergh}}]{Thorstensen}
{Thorstensen}, J.~R., {Fesen}, R.~A., \& {van den Bergh}, S. 2001, \aj, 122,
  297

\bibitem[{{Uchiyama} \& {Aharonian}(2008)}]{Uchiyama}
{Uchiyama}, Y. \& {Aharonian}, F.~A. 2008, \apjl, 677, L105

\bibitem[{{Vink}(2006)}]{Vinkreview2006}
{Vink}, J. 2006, ArXiv Astrophysics e-prints

\bibitem[{{Vink}(2008)}]{Vink2008}
---. 2008, \aap, Accepted

\bibitem[{{Vink} {et~al.}(1998){Vink}, {Bloemen}, {Kaastra}, \&
  {Bleeker}}]{Vink1998}
{Vink}, J., {Bloemen}, H., {Kaastra}, J.~S., \& {Bleeker}, J.~A.~M. 1998, \aap,
  339, 201

\bibitem[{{Vink} {et~al.}(1996){Vink}, {Kaastra}, \& {Bleeker}}]{Vink1996}
{Vink}, J., {Kaastra}, J.~S., \& {Bleeker}, J.~A.~M. 1996, \aap, 307, L41

\bibitem[{{Vink} \& {Laming}(2003)}]{Vink2003}
{Vink}, J. \& {Laming}, J.~M. 2003, \apj, 584, 758

\bibitem[{{Vink} {et~al.}(1999)}]{Vink1999}
{Vink}, J. {et~al.} 1999, \aap, 344, 289

\bibitem[{{Vink} {et~al.}(2001)}]{Vink2001}
---. 2001, \apjl, 560, L79

\bibitem[{{Willingale} {et~al.}(2002){Willingale}, {Bleeker}, {van der Heyden},
  {Kaastra}, \& {Vink}}]{Willingale2002}
{Willingale}, R., {Bleeker}, J.~A.~M., {van der Heyden}, K.~J., {Kaastra},
  J.~S., \& {Vink}, J. 2002, \aap, 381, 1039

\bibitem[{{Willingale} {et~al.}(1996){Willingale}, {West}, {Pye}, \&
  {Stewart}}]{Willingale}
{Willingale}, R., {West}, R.~G., {Pye}, J.~P., \& {Stewart}, G.~C. 1996,
  \mnras, 278, 749

\bibitem[{{Yang} {et~al.}(2007){Yang}, {Lu}, \& {Chen}}]{Yang}
{Yang}, X.-j., {Lu}, F.-j., \& {Chen}, L. 2007, ArXiv e-prints, 712

\bibitem[{{Zirakashvili} \& {Aharonian}(2007)}]{Zirakashvili}
{Zirakashvili}, V.~N. \& {Aharonian}, F. 2007, \aap, 465, 695

\end{thebibliography}
\end{document}